# Engineering Semantic Web Applications by Using Object-Oriented Paradigm

Amjad Farooq, Syed Ahsan and Abad Shah

**Abstract:** The web information resources are growing explosively in number and volume. Now to retrieve relevant data from web has become very difficult and time-consuming. Semantic Web envisions that these web resources should be developed in machine-processable way in order to handle irrelevancy and manual processing problems. Whereas, the Semantic Web is an extension of current web, in which web resources are equipped with formal semantics about their interpretation through machines. These web resources are usually contained in web applications and systems, and their formal semantics are normally represented in the form of web-ontologies. In this research paper, an object-oriented design methodology (OODM) is upgraded for developing semantic web applications. OODM has been developed for designing of web applications for the current web. This methodology is good enough to develop web applications. It also provides a systematic approach for the web applications development but it is not helpful in generating machine-pocessable content of web applications in their development. Therefore, this methodology needs to be extended. In this paper, we propose that extension in OODM. This new extended version is referred to as the semantic web object-oriented design methodology (SW-OODM).

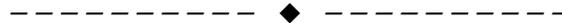

## 1. INTRODUCTION

The Web encompasses huge volume of information resources. It is extensively used by several users for retrieving their required information but most of them are unable to retrieve relevant, exact and specific information for their immediate needs. Its main reason is the way the contents of the current web information resources are stored and maintained. These contents are usually stored in the machine-readable form, but not in the machine understandable form. This way of storing and maintaining web resources is considered a root cause of irrelevancy in information retrieval. To make contents machine understandable, Tim Berners-Lee [1] gave the concept of the Semantic Web (SW). SW is an extension of the current web, and it is generally called 'a web of ontologies'; whereas the current web is called 'a web of documents containing data'. SW is an extended form of the current web, which means that SW contains both characteristics: (i) web of documents containing data and (ii) web of ontologies. Ontology *formally* provides structural knowledge of a domain and its data in the machine-understandable way. World Wide Web Consortium (W3C) has recommended SW technologies (such as Resource Description Framework Schema (RDFS) and Web Ontology Language (OWL)) for ontology formalizing

*Amjad Farooq, Syed Ahsan and Abad Shah: Computer Science and Engineering Department, University of Engineering and Technology, Lahore –Pakistan.*

Semantic web applications are fundamental requirements of SW [2], [3].

A web application, equipped with ontology, is referred as *semantic web application*. In other words, a SW application consists of semantic web pages instead of simple/ordinary web pages. A semantic web page contains content is natural language as well as in the ontological form. OODM [4] is not suitable for the development of semantic web applications. This methodology need to be extended so that they could fulfill and handle the new requirements (i.e. content generation in the ontological form) of SW applications. Note that OODM was proposed for the development of applications for the current web. OODM mainly focuses on its two development phases, i.e., (i) Analysis Phase, (ii) Design Phase. The working of these two phases is provided in an algorithmically form. This is a model-driven methodology because in both phases different formal models are used for modeling different processes. All these models are briefly described in Section 2 and their details are available in book chapter entitled "Information Modeling for Internet Applications" [4]. As we have mentioned earlier that we extend this methodology (OODM) to make it suitable for the development of SW applications and this extended version of OODM is referred to as *Semantic Web OODM (SW-OODM)*.

The remainder of this paper is organized as follows. An overview of OODM is given in Section 2. A brief comparison of current Web vs. Semantic Web is given in the same Section. This Section also gives brief description of existing methodologies for developing semantic web applications. Proposed extension i.e., SW-OODM is given in Section 3, and then it is validated in Section 4 via a case study. Finally, the paper ends with conclusion and future work in Section 5.

## 2. AREA BACKGROUND AND RELATED WORK

### 2.1 Semantic Web vs. Current Web

Semantic web may be compared with non-semantic (i.e. current web), within several parameters such as *content, conceptual perception, scope, environment* and *resource-utilization.*

*(a)* **Content**: Semantic web encompasses actual content along with its formal semantics. Here formal semantics are machine understandable content, generated in logic-based languages such as Web Ontology Language (OWL) recommended by W3C. Through formal semantics of content, computers can make inferences about the data i.e., to understand what a data resource is and how it relates to other data. In today's web there is no formal semantics of existing contents, these content are machine-readable but not machine understandable [5],[6].

*(b)* **Conceptual Perception**: Current web is just like a book, having multiple hyperlinked documents. In book scenario, index of keywords are there in each book but context in which those keywords are used is missing in the indexes. There are no formal semantic of keywords in indexes. To check which one is relevant, we have to read the corresponding pages of that book. Same is the case with current web. In semantic web this limitation will be handled via the concept of ontologies, where data is given well-defined meanings, understandable by machines [1],[6].

*(c)* **Scope**: Through literature survey [7],[8] it has been determined that inaccessible part of the web is about five hundred times larger than accessible one. It is estimated that there are billion pages of information available on the web, and only a few of them can be reached via traditional search engines. In semantic web formal semantics of data are available via ontologies, and the ontologies are the essential component of semantic web, accessible to semantic search engines.

*(d)* **Environment**: As Tim B. Lee said that semantic web is the web of ontologoies having data with formal meanings [9]. This is in contrast to current web which contains virtually boundless information in the form of documents. The semantic web, on the other hand, is about having data as well as documents that machines can process, transform, assemble, and even act on data in useful ways.

*(e)* **Resources Utilization**: There are a lot of web resources that may be very useful in our everyday activities. In current web it is difficult to locate them; because they are not annotated properly by the metadata understandable by machines. In semantic web there will a network of related resources. It will be very easy to locate and to use them in semantic web world.

Similarly there are some other criteria factors for comparison between current web and semantic web. For example, Information searching, accessing, extracting, interpreting and processing on semantic web will be more easy and efficient; Semantic web will have inference or reasoning capability; Network or communication cost will be reduce in the presence of semantic web for the reason that of relevant results; and many more - some are listed in the following table.

### 2.2 Object-Oriented Design Methodology (OODM)

OODM [2] is a model-driven methodology for the developing web applications. This methodology emphasizes on capturing and modeling the users' needs during analysis phase rather than in design phase. It makes activities more manageable and controllable. In the analysis phase of this methodology, the problem statement is studied. Information and their structure, potential users and their goals, navigation paths, and operations supported by the web application are identified. Three models are prepared from problem statement. They are information model, user navigation model and operation model. In the design phase, the presentations of information to users, user navigation paths, implementation of each operation, and user-interface elements are designed. This methodology clearly provides processing of each phase in algorithmically form.

### 2.3 Existing Methodologies for developing semantic web applications

We have reviewed the literature concerning the adaptation of existing web development methodologies. The findings disclose that, not much work has been done in this direction.

XML web engineering methodologies adaptations are proposed in the form of WEESA [10]. It generates semantic annotations by defining a mapping between the XML schemas and existing ontologies. WEESA cannot directly use domain ontologies created/reused during the web design process, but instead need to define this map ping regardless if a domain ontology was used during the design process or not. Data modeling is done twice: once in the XML schema, once in the domain ontology used.

In [11], the authors have proposed an extension of the Web Site Design Method. In this approach object chunk entities are mapped to concepts in the ontology. OOHDM [12] [13], has extended in the light of the Semantic Web technologies.

| Sr. No | Web Factors | (Non-Semantic) Web | Semantic Web |
|---|---|---|---|
| 1. | Conceptual Perception | large hyperlinked book | large interlinked database |
| 2. | Content | No formal meanings | Formally defined |
| 3. | Scope | Limited – Probably invisible web excluded | Boundless - Probably invisible web included |
| 4. | Environment | Web of documents | Web of ontologies, data and documents |
| 5. | Resources Utilization | Minimum-Normal | Maximum |
| 6. | Inference/Reasoning capability | No | Yes |
| 7. | Knowledge Management applications sport | No | Yes |
| 8. | Information searching, accessing, extracting, interpreting and processing | Difficult and time-consuming task | Easy and Efficient |
| 9. | Timeliness, accuracy, transparency of information | Less | More |
| 10. | Semantic Heterogeneity | More | Less |
| 11. | Ingredients | -Content<br>-Presentation | -Content<br>-Formal Semantics<br>-Presentation |
| 12. | Text simplification and clarification | No | Yes |

Tab 1. Semantic web characteristics

| Analysis Phase | | | |
|---|---|---|---|
| **Component** | **Input** | **Output** | **Output-detail** |
| Building Information model | Problem statement | Information model | Page-classes, multimedia attributes, and associations among page-classes |
| Building User navigation model | Problem statement<br>Information model | User navigation model | User classes, user goals,<br>User access scenarios,<br>User navigation paths |
| Building Operation model | Problem statement<br>Information model<br>User navigation model | Operation model | Operation names, input, output, and dynamic page-classes |
| Design Phase | | | |
| Building Component model | Information model | Component model | Components and components access sequence |
| Building Navigation model | Component model<br>User navigation model | Navigation model | Local navigation, instance navigation, global navigation, and menu navigation |
| Building Operation partitioning model | Operation model | Operation partitioning model | Client and server operations |
| Building User interface model | User navigation model Component model Navigation model Operation model | User interface model | User interface elements Navigation primitives User interface elements Forms user interface elements |

Tab 2: The outline of OODM

Its primitives for conceptual and navigation models have been described as DAML classes and RDFS have been used for domain vocabulary.

The HERA [13] methodology has been extended for adaptive web-based application engineering. It uses semantic web description formats for models representation.

## 3. PROPOSED METHODOLOGY

SW-OODM encompasses certain activities to formalize content and descriptive knowledge of an application domain in machine understandable and human readable and understandable formats. This methodology mainly

### 3.1 Analysis Phase

All activities of analysis phase are grouped into four models: preliminary web-ontology model, information model, user model and operation model. The activities of last three models are mostly same as for OODM. The detail of these models may be seen in Section 2 or in chapter given in book entitled "Information Modeling for Internet Applications [2].

For developing semantic web application, a new model called *preliminary web-ontology model*, is constructed in this phase. This model captures all requirements necessarily to develop a web-ontology of particular domain in order to enable contents and descriptive knowledge of that domain in machine-processable format. This model consists of different steps. Their brief descriptions are given below.

**a) Domain Vocabulary Declaration:** Domain vocabulary is the foundation of a semantic web application [15],16]. It is prepared with consensus and consultation of domain experts, ontology engineers and web engineers involved in development of application, in order to avoid semantic heterogeneity. Commonly occurring words and phrase denoting domain concepts are properly named. Attributes and verbs along with their description are also included in domain vocabulary list. All vocabulary terms are defined in a namespace, referenced by some Universal Resource Identifier (URI).

**b) Identify resources and assign them to proper groups**: In web context, a resource is any thing that has URI or can be referenced by an URI. Therefore if some concept has number of instances then all those instances are grouped in a class and that class is included in the resource-list. Multiple classes can belong to same page, and each page is represented by a URI, therefore that page is also included in the resource-list. Similar pages may be grouped into a page-class, represented by an URI; therefore that page-class should be included in the resource-list.

| Component | Input | Output-detail |
|---|---|---|
| **Preliminary web-ontology Model** | Problem statement | • domain Vocabulary<br>• list of Resources<br>• named resource-relationships<br>• resources data characteristics<br>• constraints |
| **Information Model** | Problem statement | Page-classes, multimedia attributes, and associations among page-classes |
| **User Model** | Problem statement<br>Information model | User classes, user goals,<br>User access scenarios,<br>User navigation paths |
| **Operation Model** | Problem statement<br>Information model<br>User navigation model | Operation names, input, output<br>And dynamic page-classes |

Tab 3. Analysis phase of SW-OODM

**c) Identify Axioms:** The structural knowledge about, how the resources interact with each other, may be specified in term of axioms. These are sentences written by using domain vocabulary. These represent declarative knowledge about concepts, and these are always accepted to be true without any proof. Axioms also represent semantics about behavior and properties of concepts. Usually axioms are interpreted as rules for concepts and we can drive different information about concepts from those axioms. All domain axioms are listed in this activity.

**d) Identify relationships and assign them proper names:** From axioms listed in previous step, interactions between resources are determined and proper names are assigned to each relationship from domain vocabulary produced in step a). If some one is missing, go step (a) and define them. Inverse name is also defined and listed for each relationship to make application more knowledge- enriched.

**e) Identify data-characteristics and assign them proper names:** A characteristic is a specific feature, attribute, or element used to describe a resource. Each characteristic has a specific meaning. It defines its permitted values, the types of resources it can describe, and its relationship with other characteristics. Assign proper name to each data-characteristic from domain vocabulary. If some one is missing, go step (a) and define them. Inverse name is also defined and listed for each name to make application more knowledge-enriched.

**f) Apply constraints:** A domain for a named-relationship specifies which resources are potential subjects of statements, having that named-relationship as predicate. Here the statement is the basic element of preliminary web-ontology model, it has triplet format and the predicate is one of them (i.e. subject, predicate and object). Domain of named-relationship consists of classes because classes encompass resources. There can be multiple classes in the domain of named-relationship. A range for a named-relationship specifies which resources may become objects of statements those have that named-relation as predicate. Again, there can be multiple classes in the range of a named-relationship.

**g) Verification:** Although this activity should be carried out in parallel in each above step, but after completion of preliminary web-ontology model, it should be again tested for its consistency, correctness and completeness with the help of domain experts. For consistency the defects such as, using more than one name for same resource, and an individual assignment to two mutually disjoint classes are diagnosed. In completeness testing, the defects such as omission of domain resources and the omission of relationship are diagnosed. Similarly, in correctness testing the use of incorrect relationship, aggregation & specialization of classes and cardinality of relationship are tested. Semantic heterogeneities are also determined and dissolved in this step. The constraints on domain and range values of each object property and datatype property are also verified in the testing activity.

| S No. | Component | Input | Output-detail |
|---|---|---|---|
| i | **Web-ontology Model** | Preliminary Web-ontology Model | - Triples<br>- RDF-graph<br>- OWL-description (Web-ontology Schema) |
| ii | **Component Model** | Information model | Components and components access sequence |
| iii | **Navigation Model** | Component model<br>User Model | Local navigation, instance navigation, global navigation, and menu navigation |
| iv | **Operation Partitioning Model** | Operation model | Client and server operations |
| v | **Semantic Web-page Design** | Navigation model<br>Component model<br>Operation Model<br>Web-ontology | Page Templates<br>- User-template<br>- Machine Template<br>Instantiated templates |
| vi | **User Interface Model** | User navigation model<br>Component model<br>Navigation model<br>Operation model | User interface elements<br>Navigation primitives<br>User interface elements<br>Forms user interface elements |

Tab 4. Design phase of SW-OODM

## 3.2 Design Phase

Design phase of SW-OODM works into sixes units, namely, building component model, building navigation model, building operation partitioning model, web-ontology model and semantic web-page design.

The workings of four units (ii), (iii), (vi), (vi) are already given in OODM, whereas the rest of models are discussed below:

### (i) Web-ontology model

It contains formal description of preliminary domain model produced in analysis phase and it may contain instance data as well. Both of these are represented as a set of triples, and these can be shown in the form of graph. Each resource and its instance are represented using a set of statements describing the same resource. In any triple the subject or object can be a blank node. A blank node is not a URI reference or a literal. Individual to individual relationship are also listed in triplet forms. Classes are assigned to web pages. Since the concept CLASS is not the same as it is in OO paradigm; here the class is a set of individuals, and an individual can belong to more than one, classes.

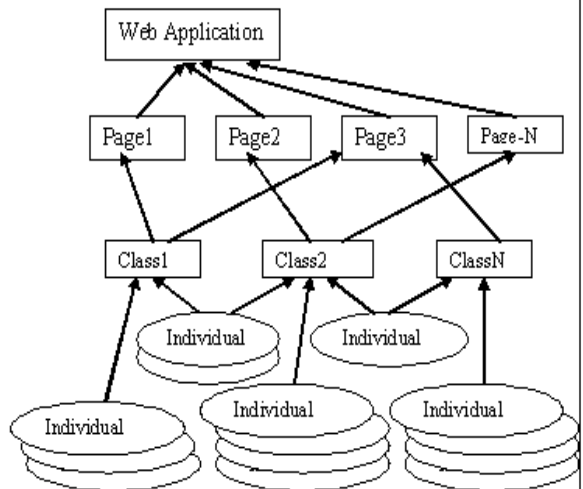

Fig 1. Relationship b/w an individual and web application via classes and pages

On completion of this phase, the web-ontology model evaluated for its conformity with domain model produced in previous phase. Furthermore it is also ensured the consistency of individuals, especially for those, belonging to more that one classes. As we know, ontology also represents the rules of integrity and inferences, i.e. x is fatherOf y; its means that age of every individual of type x must be greater than its corresponding individual of type y; so this type of validating activity is also performed on rdf model. Furthermore the logical completeness and computational completeness is also evaluated.

### (v) Semantic Web Page Design

For simple web-page design a detailed guideline is given in OODM. Now for designing semantic web-page (SWP), divide page design into sections: machine-template and user-template.

### a) Semantic web-page Template Design

- Machine-template: Template for Machine processable contents
  The machine-template section of a SWP will contain formal semantics (machine-processable) of user understandable content, stored in user-template section of same page. The contents of machine-template will be invisible for human-user. It is a text area, and it contains owl-code.
- User-template: Template for Human-user understandable contents
  Design this template according to guideline given in OODM.

### b) Semantic web-page Template Instantiation

First of all, fetch data from data store into a class-buffer as page-class instances and then generate page contents (for machine as well as human-user). The page-class controls the content consistency. Whenever the data need to be changed, the changes will be made in class-buffer and then both the machine-template and human-user template of the page will be re-populated from class-buffer automatically.

## 4. CASE STUDY

In this section, we present a case study to illustrate only those steps of the analysis and design phases of proposed methodology that focuses on the development of semantic web pages. For this purpose, we develop a semantic web application for CS&E department of a University. In this paper we have illustrated the "Research Activities Management" section of semantic web application, omitting other sections due to space limitation.

### 4.1 Analysis Phase

### (i) Preliminary web-ontology Model

a) Domain Vocabulary

There may be number of research groups, each has a URI, so these are grouped in a class so-called "ResearchGroup". A person entitled as Director directs each research grouped. There may be several researchers, having unique URIs; we have grouped them in a class called "Researcher". Each researcher may have different data-characteristics such as name, job-title, joining-date, email and mobile number; these data-characteristics are formally termed as *hasName, hasJobTitle, hasStartingDate, hasEmail,* and *hasCell* respectively. Research paper is a very common concept used in research activities. There may be several papers written by different authors, as each paper is referenced by a URI, therefore we group them in a *ResearchPaper* class. Each research paper has title, author(s), abstract and publishing-year, paper-type and body section. These characteristics are formally termed as *hasAuthor, hasTitle, hasAbstract, hasPublishingYear, hasCategory,* and *hasTextURI* respectively. Each group conducts its research into different research areas, where each area has certain title along with short description and is unique-identifier. Description and unique-identifier characteristics are formally termed as *hasDescription* and *hasId* respectively where the title is already termed as hasTitle. All instances of research area are grouped in a class, formally named as *ResearchArea*. A research paper has category that depends on the medium where it is published. It may be a published in local conference, in an international conference, in research journal or my be a chapter in a book. It has recommended that a separate class should be used for holding paper category information, and we have named that class as *paperCategory*. Each instance of paperCategory may have id, title and short description.

(b) Identify resources and assign them to proper groups

*RearchGroup*: There may be number of research groups. These are grouped in a class so-called "ResearchGroup". There may be several researchers. We have grouped them in a class called "Researcher". Paper is a very common concept used in research activities. There may be several papers written by different authors, as each paper may be referenced by a URI, therefore we group them in a *ResearchPaper* class. Each group conducts its research into different research areas, where each area has certain title along with short description and is uniquely identified by a URI; we group them in a *ResearchArea* class. A research paper may be a published in local conference, in an international conference, in research journal or a chapter in a book. Each individual of paperCategory may have id, title and short description.

(c) Identify Axioms

*A ResearchGroup has ResearchArea; A ResearchGroup has a director; A ResearchArea has deputyDirector; A director and deputyDirector are Researchers; A Researcher may be a student, faculty or a software engineer; A researcher writes ResearchPaper; A researchPaper may be a National Conference Paper; A researchPaper may be an International Conference Paper; A researchPaper may be a Journal article; A researchPaper may be chapter in some Book; A researchPaper has author(s); An author is a Researcher; A researchPaper has a Title; A researchPaper has text; A researchPaper has publishing year.*

(d) Identify relationships and assign them proper names

Relationship between ResearchGroup and Researcher is named as hasDirector. *hasDeputyDirector* is a named-relationship between ResearchArea and Researcher. *hasAuthor* is a named-relationship between researchPaper and Researcher. *hasArea is a* named-relationship that exists between ResearchGroup and ResearchArea class. Relationship between ResearchPaper and PaperCategory is named as hasCategory. Similarly, relationship between Researcher and ResearchArea is named as *hasResearchArea*. We have chosen a few relationships, by excluding others due to space limitation.

(e) Identify data-characteristics and assign them proper names

| Name | Domain Class | Range Class |
| --- | --- | --- |
| *hasId* | *ResearchGroup* *ResearchArea* *Researcher* *PaperCategory* | Number datatype |
| *hasTtile* | *ResearchGroup* *ResearchArea* *Researcher* *ResearchPaper* *PaperCategory* | String |
| *hasEmail* | *Researcher* | String |
| *hasName* | *Researcher* | String |
| *hasCell* | *Researcher* | Number |
| *has Affiliation* | *Researcher* | String |
| *hasStartingDate* | *ResearchGroup* *ResearchArea* *Researcher* | Date |
| *hasText* | *ResearchPaper* | PageURI |
| *hasPublishingYear* | *ResearchPaper* | Number |

Tab 5. Data-elements with domain and range constraints

(f) Apply constraints of named-relationships

| Name | Domain Class | Range Class |
| --- | --- | --- |
| *hasDirector* | *ResearchGroup* | *Researcher* |
| *hasDeputyDirector* | *ResearchArea* | *Researcher* |
| *has Area* | *ResearchGroup* | *ResearchArea* |
| *hasAuthor* | *ResearchPaper* | *Researcher* |

| hasCategory | ResearchPaper | PaperCategory |
| hasResearchArea | Researcher | ResearchArea |

Tab 6. Resource relationships along with domain and range constraints

*g) Validating*

With the help of domain experts we have evaluated outputs of all activities of this phase to check their completeness and correctness. Different types of tests as mentioned in validation section of specification phase in proposed section are performed before switching to next phase.

**4.2 Design Phase**

*Web-ontology Model:* Transform preliminary web-ontology model into a ontology model and that encompasses triples. We have represented the output of previous phase in triples using the format; *Predicate [Subject, Object]*

```
is_a [Author, Class]                    subClassOf [Faculty, Person]
is_a [Person, Class]                    is_a [hasTitle, DatatypeProperty]
subClassOf [Author, Person]             hasDatatype [hasTitle, String]
is_a [hasCategory, ObjectProperty]      hasDirector [ResearchGroup, Director]
hasRange [hasCategory, PaperCategory]   hasArea [ResearchGroup,
hasDomain [hasCategory, ResearchPaper]  ResearchArea]
is_a [belongToPage, ObjectProperty]     hasAuthor [ResearchPaper, Author]
is_a [hasArea, ObjectProperty]          hasCategory [ResearchPaper,
is_a [hasAuthor, ObjectProperty]        PaperCategory]
hasDomain [belongToPage,                hasResearchArea [Researcher,
HardCodedContent]                       ResearchArea]
hasDomain [Author, ResearchPaper]       is_a [hasStartingDate,
hasRange [belongToPage, PageClass]      DatatypeProperty]
is_a [Director, Class]                  hasDatatype [hasStartingDate, DATE]
is_a [Faculty, Class]                   ------
------
```

Fig 2. Sample slice of triples

We have used Protégé Version 3.2.1 ontology editor for formalization of ontology model produced in figure 3.

**b) Semantic web-page Design**

We have developed certain semantic web pages, using guidelines given in user-interface model and semantic web page design units of proposed methodology. Then we validated the formal semantics of different concepts in those semantic web pages, manually. Consider a simple web-page given in figure 5. Here a user can see a list of persons, currently working in different research areas at CS&E – UET.

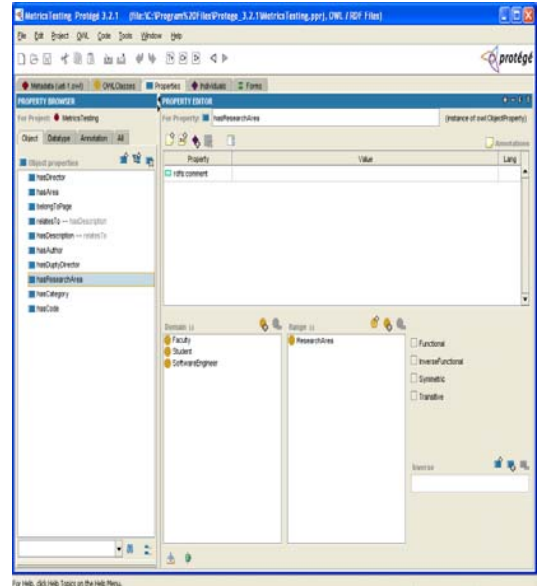

Fig 3. Protégé user-interface

Fig 4. Sample slice of ontology

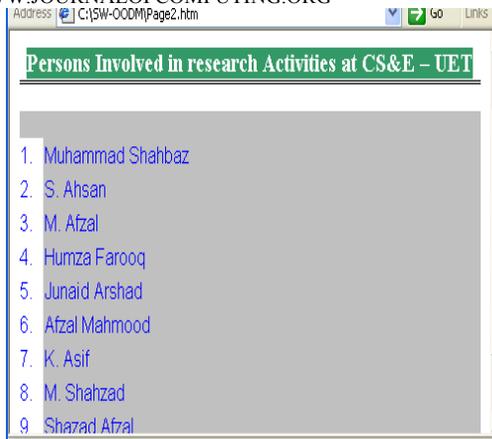

Fig 5. Non-semantic web page

In this page who is the Muhammad Shahbaz? Not known for machine; it is not formally defined because the web page is not a semantic web page. Now consider another page so-called semantic web page, given in figure 6.

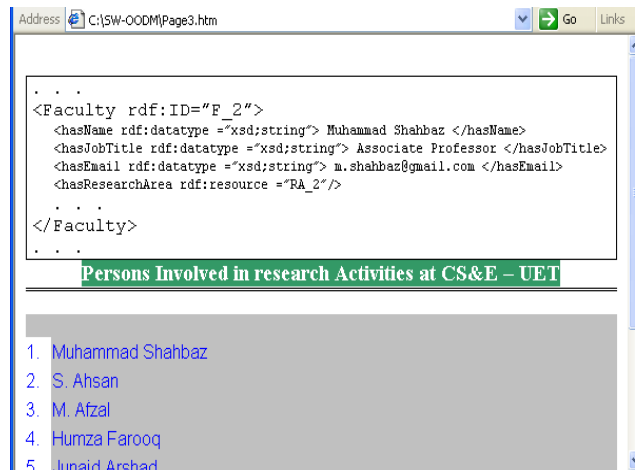

Fig 6. Semantic web-page for research personnel

Who is 'Muhammad Shahbaz'? It is known by machine; it is formally defined because the web page is a semantic web page. He is a faculty member, working as Associate Professor and is involved in "RA_2" research area. Note that boxed area (formal semantics) will be invisible for user. What's about RA_2? See another semantic web page given in figure 7. It is a research area entitled "Data Mining". What's about "RG_4"? In the formal semantics of Data Mining research area. This is a research group (see Fig 8). It includes three research areas. This research group has been started since February 06, 2004. Now this group is directed by "D_5". What's about "D_5" in the formal definition of research group "IR&SW". This is a person, having name "M. Afzal" working as Professor in the faculty of CS&E and is working as director of a research group entitled as "IR&SW" since January 25, 2006.

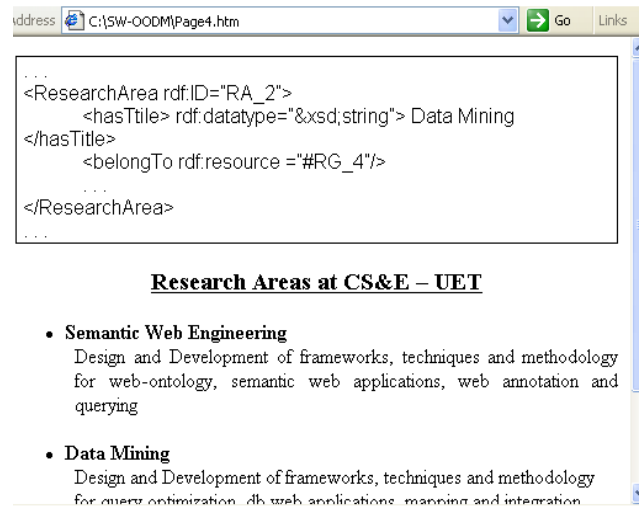

Fig 7. A semantic web-page

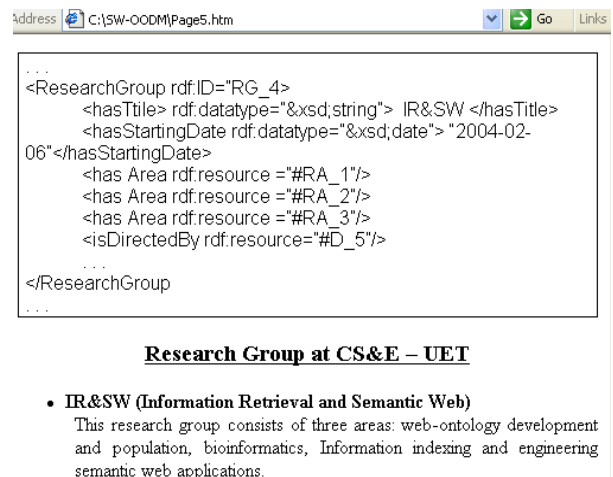

Fig 8. A semantic web page with formal semantics of research groups

In summary, Muhammad Shahbaz is an associate professor at CS&E Dept. He is working in "Data mining" research area. This area is a part of "IT & SW" research group and "M Afzal" directs this research group. All this is conformity with the definition of semantic web, i.e. resources are defined and linked in such a way, that they become understandable and processable by machine, automatically.

## 5. CONCLUSION AND FUTURE DIRECTIONS

This paper is an adaptation of OODM to develop semantic web applications. Semantic web applications consist of human readable and machine understandable contents in HTML as well as in OWL formats respectively. The adaptation encompasses three new models, named as preliminary web-ontology model,

```
. . .
<Director rdf:ID="D_5">
 <hasCode rdf:resource="#F_7"/>
 <hasJoiningDate rdf:datatype ="&xsd;date">
     2006-01-25 </hasJoiningDate>
</Director>
. . .
<Faculty rdf:ID="F_7">
    <hasName rdf:datatype ="xsd;string"> M. Afzal </hasName>
    <hasJobTitle rdf:datatype ="xsd;string"> Professor </hasJobTitle>
    <hasEmail rdf:datatype ="xsd;string"> m.afzal@hotmail.com </hasEmail>
    . . .
</Faculty>
. . .
```

Fig 9. Web-ontology code slice showing formal description of a director

ontology model and semantic web page design model. The concept *class* in web-ontology is not the same as it is used in object-oriented language. Semantic heterogeneity of individuals is automatically removed, when they are assigned to classes and their classes are grouped in sub-domains and then domains. In our future work, we have planed automate the transformation of preliminary web-ontology model into web-ontology model and to make its algorithms more efficient.

**ACKNOWLEDGEMENT**

This research work has been supported by the "Higher Education Commission of Pakistan", and the University of Engineering and Technology, Lahore.